\documentclass[sn-mathphys,Numbered]{sn-jnl}% Math and Physical Sciences Reference Style

\usepackage{graphicx}%
\usepackage{multirow}%
\usepackage{amsmath,amssymb,amsfonts}%
\usepackage{amsthm}%
\usepackage{mathrsfs}%
\usepackage[title]{appendix}%
\usepackage[usenames,dvipsnames]{xcolor}
\usepackage{textcomp}%
\usepackage{manyfoot}%
\usepackage{booktabs}%
\usepackage{algorithm}%
\usepackage{algorithmicx}%
\usepackage{algpseudocode}%
\usepackage{listings}%
\usepackage{bm}
\usepackage{siunitx}
\usepackage{caption}
\usepackage{lineno}
\newcommand{\fmrevised}[1]{{\color{black}#1}}

\begin{document}

\title[Article Title]{
Reducing Segregation in Vibrated Binary-Sized Granular Mixtures by Excessive Small Particle Introduction}

\author*[1]{Fumiaki Nakai}\email{fumiaki.nakai@ess.sci.osaka-u.ac.jp}

\author*[2]{Kiwamu Yoshii}\email{yoshii@r.phys.nagoya-u.ac.jp}
\affil*[1]{Department of Earth and Space Science, Osaka University, 1-1 Machikaneyama, Toyonaka 560-0043, Japan}

\affil*[2]{Department of Physics, Nagoya University, Furo-cho, Chikusa, Nagoya 464-8602, Japan}

\abstract{We numerically examine binary-sized granular mixtures confined between two parallel walls subjected to vertical vibration using the discrete element method.
For a size ratio of $3$ between large and small particles, we study the structure of large particles in moderately dense regimes where the combined two-dimensional packing fractions of both particle sizes exceed $1$.
When the fraction of small particles is small, segregation of the large particles occurs.
In contrast, as the fraction of small particles increases, an effective repulsion between the large particles emerges over distances greater than the large particle diameter, suppressing their segregation.
The emergence of reduction in segregation is confirmed for another size ratio, vibrational acceleration, system size, and for a case of bidisperse size distribution.
Additionally, at the size ratio of $3$, the effective repulsion induces a hexagonal phase of the large particles at packing fractions lower than in mono-component systems. This work will provide a fresh insight into granular physics, prompting further experimental and theoretical study.}


\maketitle

\section{Introduction}
Granular materials, when subjected to external excitation, exhibit pattern formation \cite{mehta2007granular, rosato2020segregation, aranson2009granular, reis2006crystallization, perera2010diffusivity,krengel2013pattern,moss2023behavioural} depending on particle characteristics such as size \cite{shinbrot2004brazil, jullien1990mechanism, jullien1992three, van2015underlying, knight1993vibration, melby2007depletion, xu2017segregation}, mass \cite{rivas2011segregation, arntz2014segregation, tripathi2013density}, and shape \cite{maione2015investigation, liu2021frictional, yuan2013segregation, borzsonyi2013granular,stannarius2022regular,lu2020particle} variations. For example, when large and small particles are introduced into a container and shaken, the larger particles tend to aggregate toward the top, a phenomenon commonly known as the Brazil nut effect \cite{shinbrot2004brazil, breu2003reversing, naylor2003air, schnautz2005horizontal, garzo2008brazil, sanders2004brazil, metzger2011all,chung2009brazil,balista2018modified}. Due to the energy dissipation during interparticle collisions, granular systems do not reach a thermodynamic equilibrium state, making it challenging to analyze their segregation phenomena using statistical mechanics \cite{edwards1989theory, edwards1999statistical, edwards2001tensorial, baule2018edwards}.
Gaining insight into the pattern formations of granular materials in fluidized states, which arise from particle characteristics, is not only crucial for advancing statistical mechanics for non-equilibrium systems but also beneficial for controlling transport and mixing processes associated with fields like chemical engineering, food engineering, and the pharmaceutical industry \cite{rosato2002perspective, beaulieu2021impact, verma2021experimental, oshitani2020dry, takada2017drag, zhang2022cfd, shenoy2015effect, yogi2021experimental, menbari2020studying, wang2016recycling}.

\begin{figure}[htbp]
    \centering
    \includegraphics[width=0.6\textwidth]{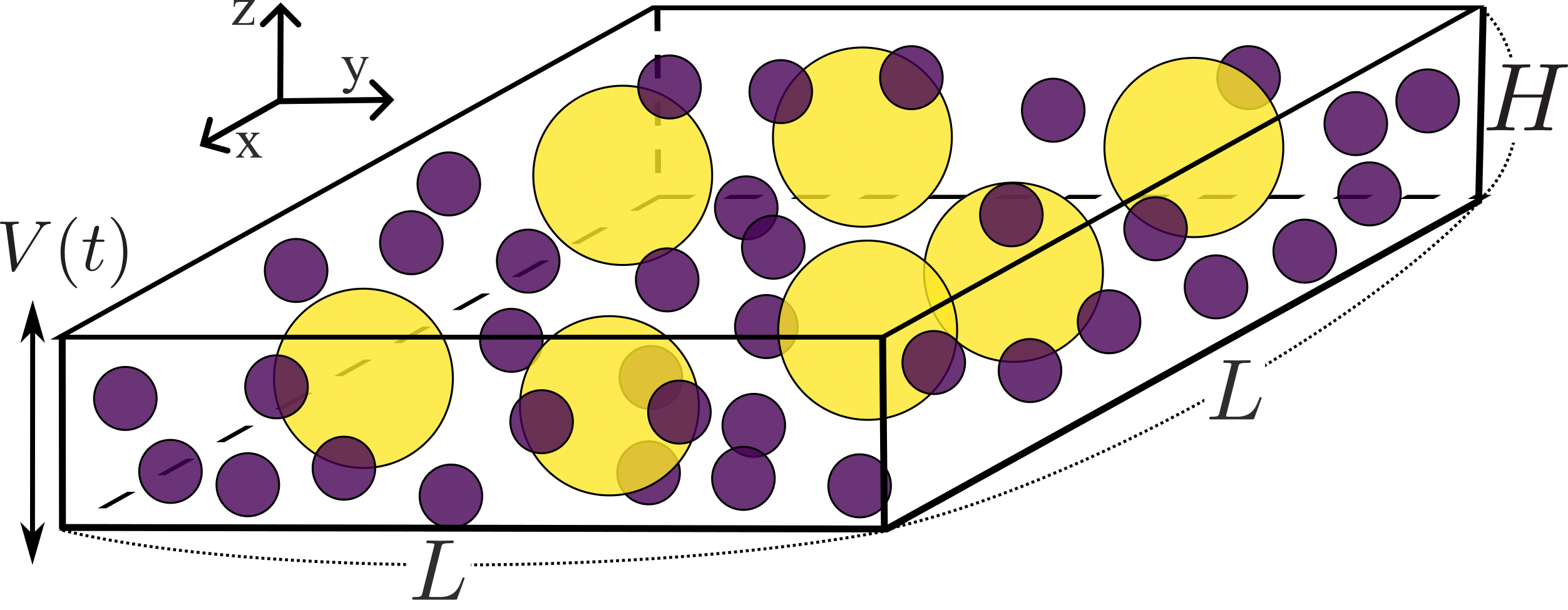}
    \caption{
    Setup targeted in this study: Binary-sized frictional granular mixtures consisting of particles with diameters of $d_l$ (yellow) and $d_s$ (purple). 
    Periodic boundaries are applied in the $xy$ directions. Vertically, the system is bounded by two parallel walls that undergo sinusoidal vibration. The box dimensions are $L$ in both horizontal directions and $H$ in height.}
    \label{fig:setup}
\end{figure}

In investigations on the pattern formation of granular materials, particles confined between two parallel walls that are vibrated in the vertical direction, as illustrated in Fig.~\ref{fig:setup}, have been intensively studied \cite{reis2006crystallization, melby2007depletion, perera2010diffusivity, plati2024quasi, rivas2011segregation, prevost2004nonequilibrium, olafsen2005two, brito2020energy, brey2014memory, safford2009structure, puglisi2012structure, wu2005structural, marschall2020depletion, bordallo2009effective, velazquez2016effective}. Such a setup ensures uniform energy input to each particle from the walls, preventing the emergence of complex inhomogeneities in granular temperature. Additionally, the ability to experimentally observe the position of each particle offers an advantage for validating theoretical and simulation studies. Despite the simplicity of this configuration, granular particles exhibit various structures depending on their mass \cite{rivas2011segregation}, size \cite{melby2007depletion, rivas2011segregation}, shape \cite{galanis2010depletion, sykes2009self, narayan2006nonequilibrium, muller2015ordering, gonzalez2017clustering, galanis2006spontaneous, galanis2010nematic}, and packing fraction \cite{reis2006crystallization, velazquez2016effective}, including gaseous states \cite{geminard2004pressure}, aggregated structures \cite{melby2007depletion, galanis2010depletion}, quasi-crystalline phase \cite{plati2024quasi}, and hexagonal phases \cite{komatsu2015roles, reis2006crystallization, wu2005structural,moss2023behavioural}. For instance, a mixture of large and small particles results in aggregation of the larger particles, manifesting a structure reminiscent of phase separation \cite{melby2007depletion, rivas2011segregation}. A similar phenomenon occurs in mixtures of rod-shaped and spherical particles \cite{galanis2010depletion}. It is emphasized that this aggregation phenomenon emerges even in the absence of attractive interactions between granular particles. 
This study targets such structures of binary-sized granular systems confined between parallel walls, as depicted in Fig.~\ref{fig:setup}.

In previous research on binary-sized granular mixtures confined between narrow parallel walls \cite{melby2007depletion, rivas2011segregation, plati2024quasi, galanis2010depletion} (also called quasi-two-dimensional systems), the structure of the large particle is examined, changing the two-dimensional packing fractions of large and small particles defined as
\begin{equation}
    \phi_{l,s}^{\text{2D}}=\frac{\pi N_{l,s}d_{l,s}^2}{4L^2}
\end{equation}
where $N_{l,s}$ denote the numbers of large and small particles, $d_{l,s}$ the diamters of large and small particles, and $L$ means the side length of the box.
So far, the study has been explored only in the cases where the combined two-dimensional packing fraction of large and small particles is less than unity, $\phi_{l}^{\text{2D}}+\phi_{s}^{\text{2D}}<1$.
Within such fraction regimes, larger particles often tend to segregate\cite{melby2007depletion}.
Here, the packing fraction for the confined binary-sized granular mixture can exceed unity since smaller particles can overlap in the vertical direction. Yet, the structural properties of these moderately dense regimes remain unexamined.
In this research, we numerically investigate the structure of the binary-sized granular system in such moderately dense regimes using the discrete element method \cite{cundall1979discrete, brilliantov2004kinetic, poschel2005computational,luding2008cohesive}.
Within the simulated size ratios, vibrational accelerations, system sizes, and distribution types (binary-size and bi-disperse), we discover that as the packing fraction of small particles increases, an effective repulsive force, longer than the diameter of the large particles, manifests between the larger particles, mitigating their segregation.
At the size ratio $3$, due to this effective repulsion, larger particles exhibit hexagonal phases at packing fractions lower than that seen in a monodisperse system \cite{reis2006crystallization, wu2005structural}.
The phenomenon of reducing segregation by introducing an excess of small particles has not been reported so far, providing a fresh insight into granular physics.
Furthermore, this finding could benefit fields like chemical engineering, food science, and pharmaceutical engineering, where granular segregation might compromise product quality or inflate costs for transportation and mixing processes.

\section{Method}
\label{sec:method}

\begin{table}[htbp]
\caption{Parameters for the DEM simulation. \label{table:parameters}}
\begin{tabular}{l|l|l}
\hline
Parameters & Symbols & Values \\ \hline
Mass density & $\rho$ & \SI{1000}{\kg\per\m^3}\\
Diameter of large particle & $d_l$ & \SI{3}{\mm}\\
Diameter of small particle & $d_s$ & \SI{0.75}{\mm} or \SI{1}{\mm}\\
Young's modulus & $E$ & \SI{3}{\GPa}\\
Poisson's ratio & $\nu$ & 0.34\\
Restitution coefficient & $e$ & 0.7\\
Coulomb friction coefficient & $\mu$ & 0.38\\
Amplitude of oscillation & $A$ & \SI{0.31}{\mm}\\
angular frequency of oscillation & $\Omega$ & \SI{890}{\radian\per\s}\\
Side length of box & $L$ & \SI{50}{mm} or \SI{100}{mm}\\
Hight of box & $H$ & \SI{3.5}{mm}\\
Number of large/small particles & $N_{l,s}$ & \\
2D packing fraction of large/small particle & $\phi^{\text{2D}}_{l,s}$ & $\pi N_{l,s}d^2_{l,s}/4L^2$\\
3D packing fraction of large/small particle & $\phi^{3D}_{l,s}$ & $\pi N_{l,s}d^3_{l,s}/6L^2H$\\
\hline
\end{tabular}
\end{table}

In this study, as displayed in Fig.~\ref{fig:setup}, the binary-sized frictional granular mixtures confined between two parallel walls are numerically investigated using the discrete element method \cite{cundall1979discrete, brilliantov2004kinetic, poschel2005computational,luding2008cohesive}.
The numbers of large and small particles are $N_l$ and $N_s$, respectively.
The diameters of large and small particles are set to be $d_l=\SI{3}{\mm}$ or $\SI{4}{\mm}$ and $d_s=\SI{1}{\mm}$, respectively.
Both particle types have a uniform mass density, set as $\rho=\SI{1000}{\kg\per\m^3}$.
The masses of large and small particles are given as $m_{l,s}=\pi \rho d_{l,s}^3/6$.
The diagonal components of the moments of inertia of large and small particles $\bm{I}_{l,s}$ are $m_{l,s} d_{l,s}^2/10$, respectively, with the other components being $0$.
The periodic boundary conditions are introduced to the $x$ and $y$ directions.
The side lengths of the box are $L=\SI{100}{\mm}$, respectively.
The height of the box is set to be $H=\SI{3.5}{\mm}$ ($H/d_L\simeq 1.17$).
When the box height is smaller than the large particle diameter, $H\le d_L$, the large particles do not almost move due to the friction between the large particle and the wall. For the large height case $H\gg d_L$, the tractable features of the confined setup, such as homogeneous injection of energy and experimentally observable configuration, will be diminished.
The gravitational acceleration $g=\SI{9.8}{\m\per\s^2}$ is imposed on all particles with $z$-axis.
The upper and bottom walls, considered as infinitely massive and large frictional spheres, act as flat walls.
These walls undergo vertical vibrations following the sinusoidal function $V(t)=A\sin(\Omega t)$, where $A=\SI{0.31}{\mm}$ and $\Omega=\SI{890}{\radian\per\s}$.
Then, the acceleration of the wall vibration is sufficiently larger than that of the gravity: $A\Omega^2/g\simeq 25$.
The time development of the $i$-th particle's position and angular velocity, $\bm{r}_i$ and $\bm{\omega}_i$, obey Newton's equation of motion:

\begin{align}
    m_i \Ddot{\bm{r}}_i &= 
    \sum_{i \neq j} (F_{ij}^{(n)}\bm{n}_{ij} + F_{ij}^{(t)}\bm{t}_{ij}) \Theta(d_{ij} - r_{ij})
    - m_i g\bm{e}_z\\
    \bm{I}_i \dot{\bm{\omega}_i} &=
    \sum_{i \neq j}
    (\bm{l}_{ij}\times \bm{t}_{ij})F_{ij}^{(t)}
\end{align}
where $F_{ij}^{(n)}$ and $F_{ij}^{(t)}$ are, respectively, the normal and tangential forces between $i$-th and $j$-th particles, $m_i$ the particle mass, $\bm{I}_i$ the inertia matrix, $d_{ij}$ the sum of the $i$-th and $j$-th particle radii, $r_{ij}$ the relative distance between the $i$-th and $j$-th particles, and $\bm{e}_z$ the unit vector along the $z$-axis.
$\bm{n}_{ij}$ and $\bm{t}_{ij}$ denote the unit vectors of the normal and tangential parts of the force between the $i$-th and $j$-th particles, $\bm{l}_{ij}$ the vector from the center of the $i$-th particle to the contact point with the $j$-th particle.
$\Theta(x)$ is the Heaviside step function, which is defined as $\Theta(x)=1$ for $x>1$ and $\Theta(x)=0$ otherwise.
In this study, we employ the Hertz/Mindlin/Tsuji contact force model \cite{tsuji1992lagrangian} for the interparticle interaction.
The normal force interaction is given by:
\begin{align}
    F^{(n)}_{ij} = k^{(n)} R_{\rm eff}^{1/2} \delta_{ij}^{3/2} - \eta^{(n)} \bm{v}_{ij}\cdot \bm{n}_{ij}.
\end{align}
Here $k^{(n)}$ represents the spring constant, $\eta^{(n)}$ viscous constant, $d_i$ the diameter of the $i$-th particle, $R_{\rm eff} = (d_i d_j)/2(d_i + d_j)$ the effective radius,
$\delta_{ij} = d_{ij} - r_{ij}$ the overlap length, and $\bm{v}_{ij}=\dot{\bm{r}}_i-\dot{\bm{r}}_j$ denotes the relative velocity.
The tangential force $F_{ij}^{(t)}$ is provided as 
\begin{align}
    F_{ij}^{(t)} = 
    -\min(|k^{(t)} R_{\rm eff}^{1/2} \delta_{ij}^{1/2}
    \bm{\xi}_{ij}^{(t)} - \eta^{(t)} \bm{v}^{(t)}_{ij}|, |\mu  F^{(n)}_{ij}|),
\end{align}
where $k^{(t)}$, $\eta^{(t)}$, and $\mu$ denote the tangential spring constant, tangential viscous constant, and friction coefficient, respectively.
The tangential velocity $\bm{v}_{ij}^{(t)}$ is defined as
\begin{equation}
\bm{v}_{ij}^{(t)} = \bm{v}_{ij} - \bm{v}_{ij}\cdot \bm{n}_{ij}\bm{n}_{ij}  - \frac{1}{2}(d_i\bm{\omega}_i + d_j\bm{\omega}_j) \times \bm{n}_{ij},
\label{eq:tangential_velocity}
\end{equation}
and $\bm{\xi}_{ij}^{(t)}$ is the tangential displacement between the $i$th and $j$th particles defined as $\bm{\xi}_{ij}^{(t)}= \int_{\tau}d\tau\bm{v}_{ij}^{(t)}(\tau)$, where $\tau$ is the contact duration between the $i$-th and $j$-th particles.
Particle-wall interaction is modeled by considering the wall as an infinitely large and massive $j$-th particle.
\fmrevised{The technical details of the simulation follow the prior work \cite{luding2008cohesive}.}
The initial arrangements of the granular particles are generated randomly to prevent overlaps. Using a time step size of $\SI{3e-7}{\s}$, we performed calculations for $10^9$ steps ($\SI{300}{\s}$) to achieve a steady state. Subsequently, main computations are also conducted for additional $10^9$ steps ($\SI{300}{s}$) to collect the data.
In the setting above, we investigate the structural property, varying 2D large and small particle packing fractions, denoted as $\phi^{\text{2D}}_{l,s}=N_{l,s}\pi d_{l,s}^2/4L^2$, respectively.
All the numerical simulations are undertaken using \textit{LAMMPS} (the open-source molecular dynamics program from Sandia National Laboratories)\cite{thompson2022lammps}.
We perform calculations using the pair style “Hertz/material,” selecting "Mindlin" for tangential force and "Tsuji" for damping force in the interaction model.
The contact force model provides $k^{(n)}$ and $k^{(t)}$ as functions of the Young's modulus $E$ and the Poisson's ratio $\nu$:
\begin{align}
    k^{(n)}&=\frac{4E}{3(1-\nu^2)}\\
    k^{(t)}&=\frac{4E}{(1+\nu)(2-\nu)}.
\end{align}
Here, we set $E=\SI{3}{\GPa}$ and $\nu=0.34$, corresponding to parameters for polystyrene \cite{chou2012discrete}.
The normal and tangential damping constants are given by:
\begin{equation}
    \eta^{(n)}=\eta^{(t)}=\alpha (m_{\rm eff} k^{(n)})^{1/2}(R_{\rm{eff}}\delta_{ij})^{1/4}
    \label{eq:damping_coefficient}
\end{equation}
where $m_{\rm eff} = m_im_j/(m_i+m_j)$ is the effective mass, and $\alpha$ the parameter relating approximately to the restitution coefficient $e$ as \cite{marshall2009discrete}
\begin{equation}
\begin{split}
    \alpha=&1.2728-4.2783e+11.087e^2-22.348e^3+27.467e^4-18.022e^5+4.8218e^6,
\end{split}
\end{equation}
We set the restitution coefficient at $e=0.7$.
Table~\ref{table:parameters} lists the parameters of the current system.

Here, when the total 2D granular packing fraction $\phi^{\text{2D}}_l+\phi^{\text{2D}}_s$ is below unity, prolonged computations can result in a complete absorbing state with respect to horizontal motions \cite{neel2014dynamics}.
Alternatively, if computations begin from an initial state without vertical overlap, the configuration on the xy-plane remains permanently unchanged.
To avoid such absorbing states, we restrict our attention to the regime where $\phi^{\text{2D}}_l+\phi^{\text{2D}}_s\ge 1$.
Additionally, when the total 3D packing fraction, given by $\phi^{3D}_l+\phi^{3D}_s$, where $\phi^{3D}_{l,s}=\pi N_{l,s}d_{l,s}^3/6L^2H$, is exceedingly large, all particles jam and drift collectively \cite{plati2022collective, scalliet2015cages, plati2020slow, plati2019dynamical}, and the initial configuration also remains almost unchanged.
In our setup, this drift behavior is predominantly evident when the total 3D packing fraction $\phi^{3D}_l+\phi^{3D}_s$ surpasses approximately $0.5$.
As the drift behavior falls outside the scope of our study, we concentrate on regimes within $\phi^{3D}_l+\phi^{3D}_s<0.5$.
Consequently, this study focuses on the regime within $\phi^{\text{2D}}_l+\phi^{\text{2D}}_s\ge 1$ (which corresponds to $\phi^{3D}_l+\phi^{3D}_s\gtrsim 0.19$ when the size ratio is $3$ in the current setup) and $\phi^{3D}_l+\phi^{3D}_s<0.5$. In this moderately dense regime, the small particles behave like a liquid; they are not uniformly distributed as in a gas state, nor are they in a jammed state.

\section{Results}

\begin{figure}[htbp]
    \centering
    \includegraphics[width=1.\columnwidth]{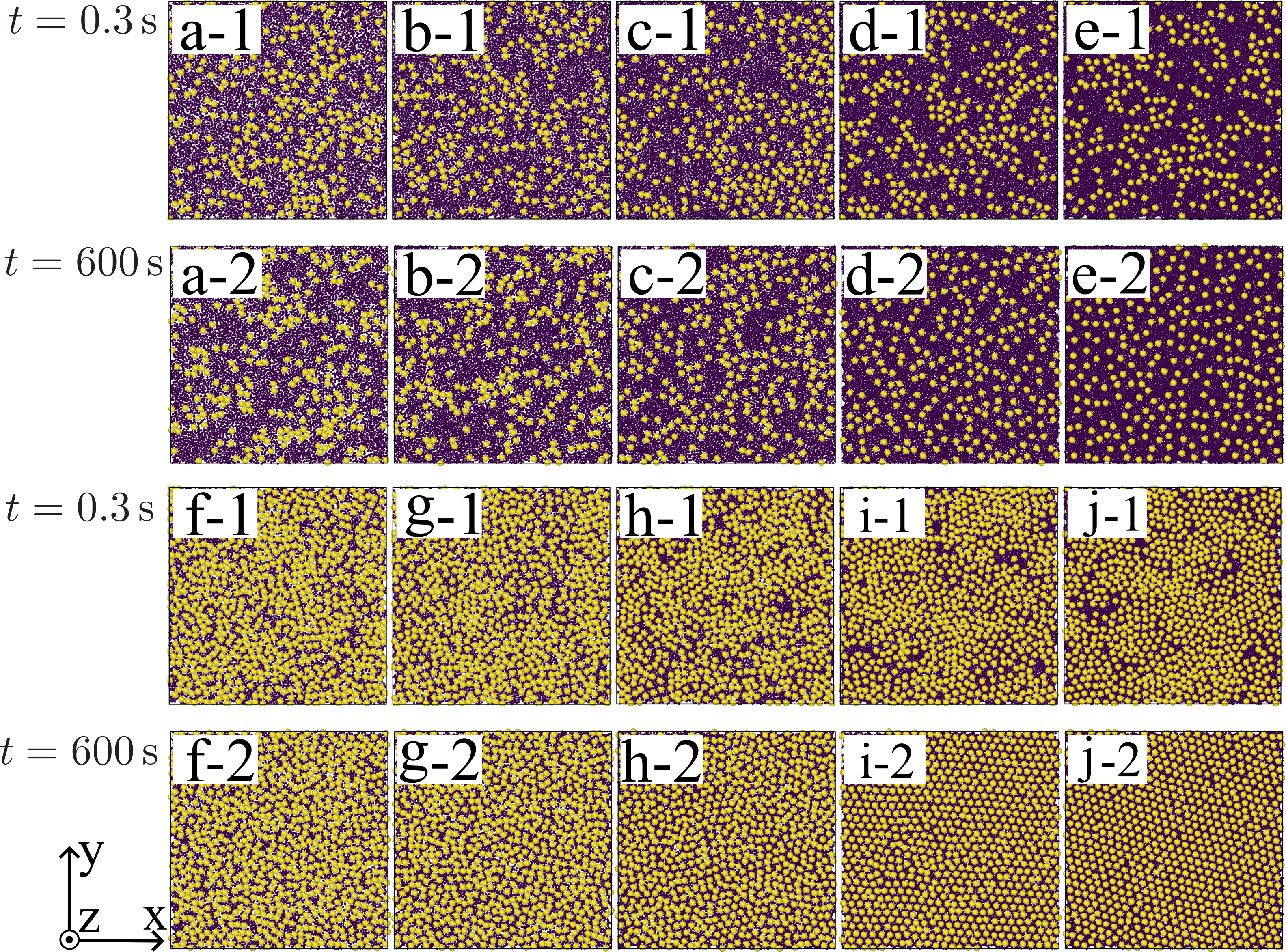}
    \caption{
    \fmrevised{Snapshots of the system with large particle diameter $d_l=\SI{3}{mm}$, small particle diameter $d_s=\SI{1}{mm}$, and system length $L=\SI{100}{mm}$, obtained through discrete element method simulations. The packing fractions of large and small particles ($\phi^{\text{2D}}_l$, $\phi^{\text{2D}}_s$) are: (a-1,2) (0.2, 0.8), (b-1,2) (0.2, 0.9), (c-1,2) (0.2, 1.0), (d-1,2) (0.2, 1.2), (e-1,2) (0.2, 1.4), (f-1,2) (0.55, 0.45), (g-1,2) (0.55, 0.5), (h-1,2) (0.55, 0.6), (i-1,2) (0.55, 0.7), and (j-1,2) (0.55, 0.8). Subscripts 1 and 2 denote snapshots taken immediately after vibration $t=\SI{0.3}{s}$ and after the long-time vibrations $t=\SI{600}{s}$, respectively. All snapshots are projected onto the $xy$-plane for clarity.}
    \label{fig:snapshot}}
\end{figure}

\fmrevised{The representative snapshots at the time immediately after vibration ($t=\SI{0.3}{s}$) and after the long-time vibrations ($t=\SI{600}{s}$) with various packing fractions $\phi_l^{\text{2D}}$ and $\phi_s^{\text{2D}}$ are presented in Fig.~\ref{fig:snapshot}. $\phi_l^{\text{2D}}$ and $\phi_s^{\text{2D}}$ are varied within the regimes $\phi^{\text{2D}}_s + \phi^{\text{2D}}_l \ge 1$ and $\phi^{3D}_s + \phi^{3D}_l \le 0.5$, as explained in Sec.~\ref{sec:method}.
When $\phi_l^{\text{2D}}=0.2$ after the long-time vibrations, weak segregation seems to be observed when the small particle fraction is low ($\phi^{\text{2D}}_s=0.2$). This weak segregation diminishes as $\phi^{\text{2D}}_s$ increases.}
When the packing fraction of small particles is large ($\phi^{\text{2D}}_s = 1.2$ or $1.4$), the long-time vibrations lead to almost no contact between the large particles.
Similarly, for $\phi^{\text{2D}}_l = 0.55$, the contacts between large particles decreases as $\phi^{\text{2D}}_s$ increases.
Moreover, for $\phi^{\text{2D}}_s = 0.7$ and $0.8$, a hexagonal phase appears to emerge. It is noteworthy that this packing fraction $\phi^{\text{2D}}_l = 0.55$ is lower than that at which a single-component case expresses the hexagonal phase ($\phi^{\text{2D}}_l \simeq 0.65$) \cite{reis2006crystallization}. To the best of the authors' knowledge, there have been no reports on behavior where excess small particles suppress segregation, and a hexagonal phase emerges at a lower $\phi_l$ than in the mono-component case.
\fmrevised{These results are insensitive to the system size and vibrational accelerations, provided $A\Omega^2/g\gg 1$, as presented in Figs.~\ref{fig:snapshot_acceleration} and \ref{fig:snapshot_systemsize} in Appendix.}

\begin{figure}[htbp]
    \centering
    \includegraphics[width=0.55\textwidth]{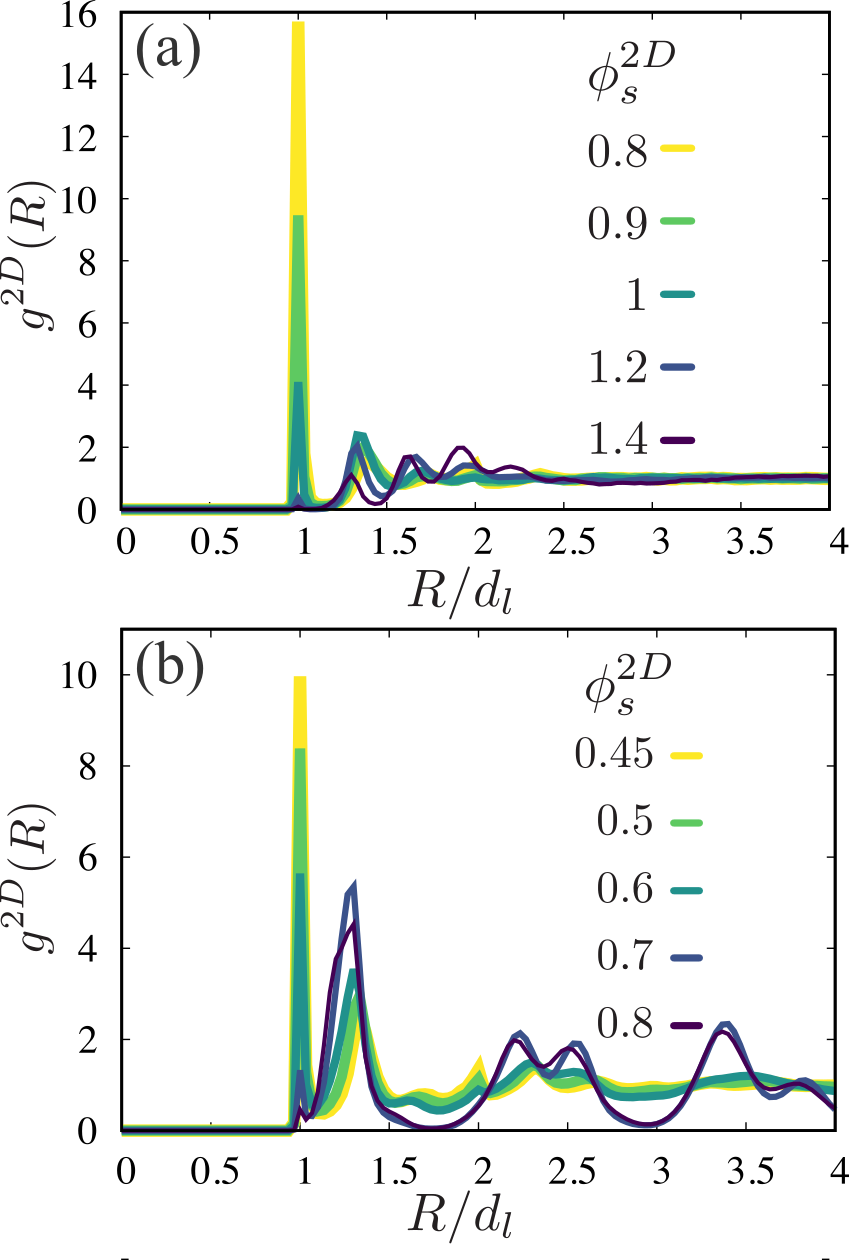}
    \caption{Radial distribution function of large particles when $d_l=\SI{3}{\mm}$ and $d_s=\SI{1}{\mm}$ for various $\phi^{\text{2D}}_s$. The horizontal axis is scaled by $d_l$.
    Representative cases are shown for (a) $\phi^{\text{2D}}_l=0.2$ and (b) $\phi^{\text{2D}}_l=0.55$.
    These data have been time-averaged over the main simulation period ($\SI{300}{s}<t<\SI{600}{s}$).\label{fig:rdf}}
\end{figure}

To characterize the arrangement of particles shown in Fig.~\ref{fig:snapshot}, we analyze the 2D radial distribution function (RDF) of large particles, defined as follows \cite{hansen2013theory}:
\begin{equation}
g^{\text{2D}}(R)=\frac{L^2}{2\pi R N_l^2}
\left\langle
\sum_i^{N_l}\sum_{j(j\ne i)}^{N_l}
\delta(R-|\bm{R}_{ij}|) \right\rangle
\label{eq:rdf}
\end{equation}
where $\delta$ is the Dirac delta function, and $\langle\cdots\rangle$ denotes the time-average.
$\bm{R}_{ij}$ represents the relative position projected onto the x-y plane between the $i$th and $j$th particles, considering the periodic boundary conditions.
Fig.~\ref{fig:rdf} displays the RDF obtained by varying $\phi^{\text{2D}}_s$ for $\phi^{\text{2D}}_l=0.2$ and $0.55$.
The horizontal axis is scaled with the diameter of the large particles, $d_l$.
For $\phi^{\text{2D}}_l=0.2$, a pronounced peak is observed at $R/d_l=1$ when $\phi^{\text{2D}}_s$ is small.
This reflects the tendency of the large particles to segregate \fmrevised{weakly}, as shown in Fig.~\ref{fig:snapshot}(a-2) and (b-2).
If the granular particles were uniformly placed, the peak value of the RDF would not be as large as our results, as calculated in 2D hard-disc systems in an equilibrium state \cite{truskett1998structural, trokhymchuk2005hard}.
As $\phi^{\text{2D}}_s$ increases, the peak at $R/d_l=1$ diminishes, while the peak at $R/d_l >1$ intensifies.
In other words, with the increase in $\phi^{\text{2D}}_s$, it can be inferred that an effective repulsion longer than $d_l$ manifests between large particles.
This trend is also noticeable for $\phi^{\text{2D}}_l=0.55$.
Furthermore, for $\phi^{\text{2D}}_l=0.55$, in alignment with the hexagonal phase observed in Fig.~\ref{fig:snapshot}(i-2) and (j-2) ($\phi^{\text{2D}}_s=0.7$ and $0.8$), the RDF does not converge to 1 even in the large $R/d_l$ region, indicating the emergence of long-ranged structures larger than the particle diameter $d_l$.
\fmrevised{These results of RDF are largely insensitive to changes in system size and vibrational accelerations, provided $A\Omega^2/g\gg 1$, as shown in Figs.~\ref{fig:rdf_acceleration} and ~\ref{fig:rdf_systemsize}.}

\begin{figure}[htbp]
    \centering
    \includegraphics[width=1.\columnwidth]{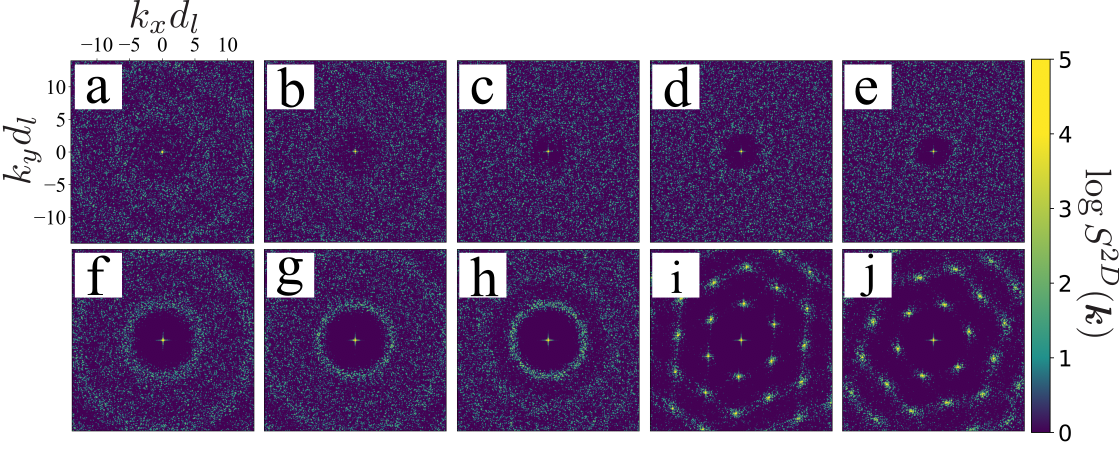}
    \caption{
    Structure factors corresponding to Fig.~\ref{fig:snapshot} at time $t=\SI{600}{s}$, as defined by Eq.~\eqref{eq:sk}, for various packing fractions of large and small granular particles. The pairs ($\phi^{\text{2D}}_l$, $\phi^{\text{2D}}_s$) are: (a) (0.2, 0.8), (b) (0.2, 0.9), (c) (0.2, 1.0), (d) (0.2, 1.2), (e) (0.2, 1.4), (f) (0.55, 0.45), (g) (0.55, 0.5), (h) (0.55, 0.6), (i) (0.55, 0.7), and (j) (0.55, 0.8).\label{fig:structure_factor}}
\end{figure}

To characterize the long-range structure of the large particles, we analyze the two-dimensional structure factor defined by the following equation \cite{hansen2013theory}:
\begin{equation}
S^{\text{2D}}(\bm{k})=\frac{1}{N_l}
\sum_{i}^{N_l} \sum_{j}^{N_l}\exp[-i\bm{k}\cdot \bm{R}_{ij}]
\label{eq:sk}
\end{equation}
where $\bm{k}=(k_x,k_y)$ denotes the wave vector.
Fig.~\ref{fig:structure_factor} shows the structure factors corresponding to the parameters in Fig.~\ref{fig:snapshot} \fmrevised{after the long-time vibrations ($t=\SI{600}{s}$)}.
No significant changes are observed from Fig.~\ref{fig:structure_factor}(a) to (e).
\fmrevised{Specifically, while the segregation of large particles decreases, as shown in Fig.~\ref{fig:snapshot}(a-2) to (e-2), no symmetric order such as a hexagonal structure appears.}
\fmrevised{This indicates that the reduction in segregation is not a consequence of the ordering of the large particles.}
In contrast to the case of small $\phi_l^{\text{2D}}$, at large $\phi_l^{\text{2D}}$ values, the increase in $\phi^{\text{2D}}_s$ leads to a structural transition from an isotropic liquid phase to a hexagonal phase, as demonstrated in Fig.~\ref{fig:snapshot}(f-2) to (j-2).
\fmrevised{Similar results for $S^{\text{2D}}(\bm{k})$ are obtained for different system sizes and vibrational accelerations when $A\Omega^2/g\gg 1$, as shown in Figs.~\ref{fig:structure_factor_acceleration} and \ref{fig:structure_factor_systemsize}.}

\begin{figure}[htbp]
    \centering
    \includegraphics[width=0.55\textwidth]{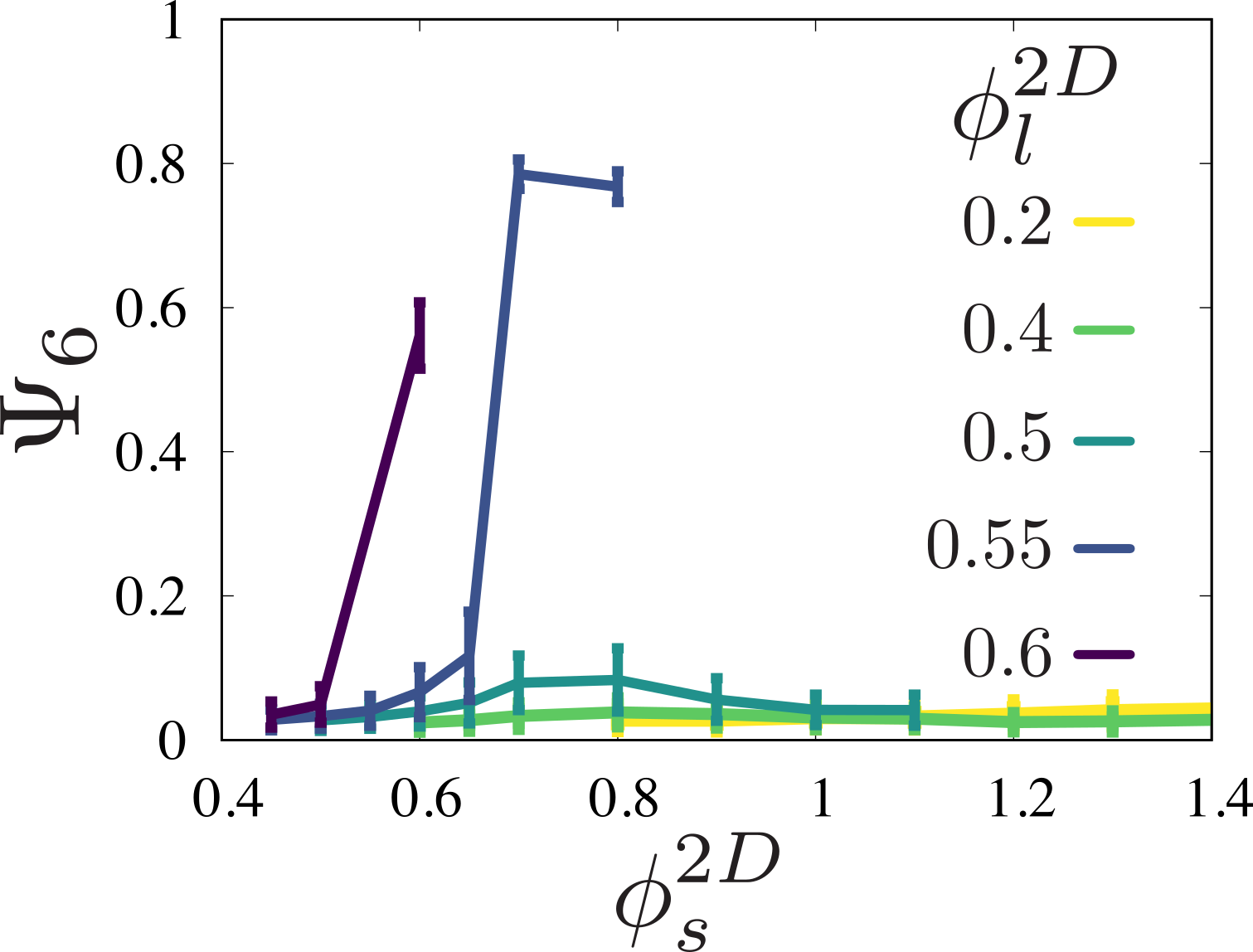}
    \caption{
    Hexatic order parameters when $d_l=\SI{3}{\mm}$ and $d_s=\SI{1}{\mm}$, defined in Eq.~\eqref{eq:order_parameter}, are plotted against $\phi^{\text{2D}}_s$ for various $\phi^{\text{2D}}_l$ values.
    The nearest neighbors are determined using Voronoi tessellation. Data are time-averaged over the main calculation period from $t = \SI{300}{s}$ to $\SI{600}{s}$.
    \label{fig:order_parameter}}
\end{figure}

To analyze the hexatic order, we calculate the global order parameter defined by the following equation \cite{reis2006crystallization}:
\begin{equation}
\Psi_6=
\left\langle
\frac{1}{N_l} \left|\sum_{i=1}^{N_l} \frac{1}{n_i} \sum_{j=1}^{n_i} \exp(i6\theta_{ij})\right|
\right\rangle.
\label{eq:order_parameter}
\end{equation}
Here, $n_i$ is the number of the nearest neighbors, determined based on Voronoi tessellation, around the $i$-th particle, $\theta_{ij}$ is the angle between the vector $\bm{R}_{ij}$ and x-axis (any fixed axis can be arbitrarily chosen).
Eq.~\eqref{eq:order_parameter} characterizes the degree of the hexagonal phase of the system, where a completely hexagonal arrangement corresponds to $\Psi_6 =1$, and a disordered structure results in $\Psi_6 \ll 1$.
Fig.~\ref{fig:order_parameter} displays $\Psi_6$ as a function of $\phi^{\text{2D}}_s$ with various $\phi^{\text{2D}}_l$, including error bars. The data are obtained as the time average within the main simulation.
For $\phi^{\text{2D}}_l<0.55$, $\Psi_6$ is nearly zero for the simulated density regime.
Meanwhile, as $\phi^{\text{2D}}_l$ is greater than or approximately equal to $0.55$, the hexatic order parameter distinctly deviates from zero for large $\phi^{\text{2D}}_s$.
This result quantitatively indicates that, upon introducing the small particles, the large particles exhibit a hexagonal phase at $\phi^{\text{2D}}_l$ values smaller than 0.65, where a monocomponent system would form a hexagonal phase \cite{reis2006crystallization}.

\begin{figure}
    \centering
    \includegraphics[width=0.8\textwidth]{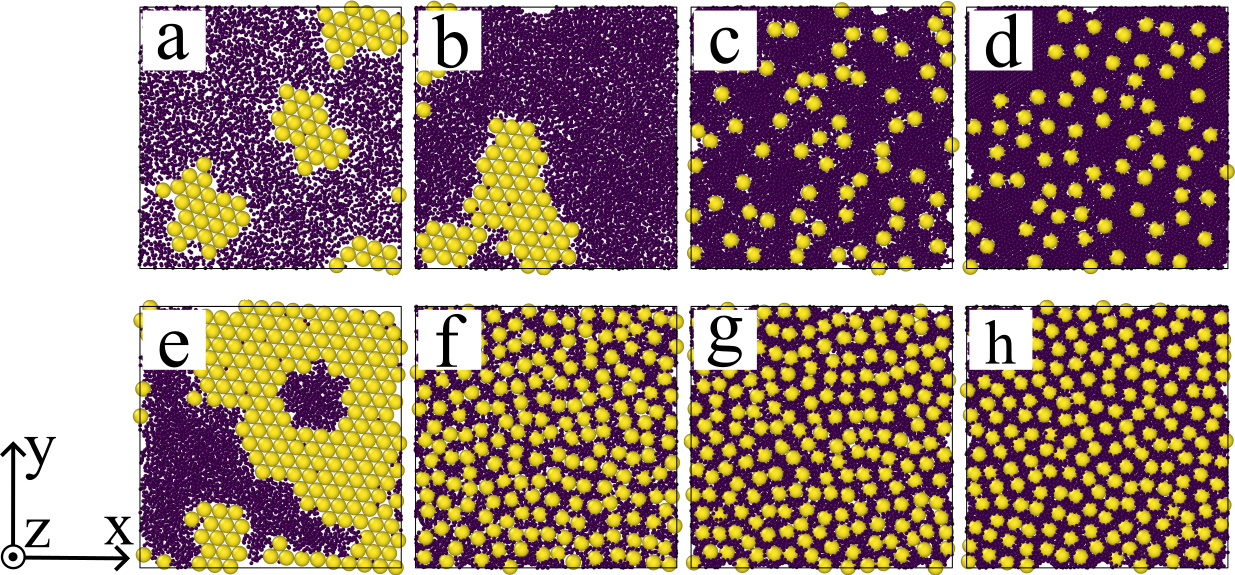}
    \caption{
    \fmrevised{
    Snapshots for the binary mixture when the size ratio is $4$ ($d_l=\SI{3}{mm}$ and $d_s=\SI{0.75}{mm}$) at the box length $L=\SI{50}{mm}$.
    The packing fractions of large and small particles $\phi^{\text{2D}}_l$, $\phi^{\text{2D}}_s$ are varied as (a)(0.2, 0.8), (b)(0.2, 1.2), (c)(0.2, 1.6), (d)(0.2, 2.0), (e)(0.55, 0.6), (f)(0.55, 0.8), (g)(0.55, 1), and (h)(0.55, 1.2). All snapshots are obtained after the long-time vibrations: $t=\SI{600}{\s}$.}}
    \label{fig:snapshot_sizeratio}
\end{figure}

\begin{figure}
    \centering
    \includegraphics[width=0.55\textwidth]{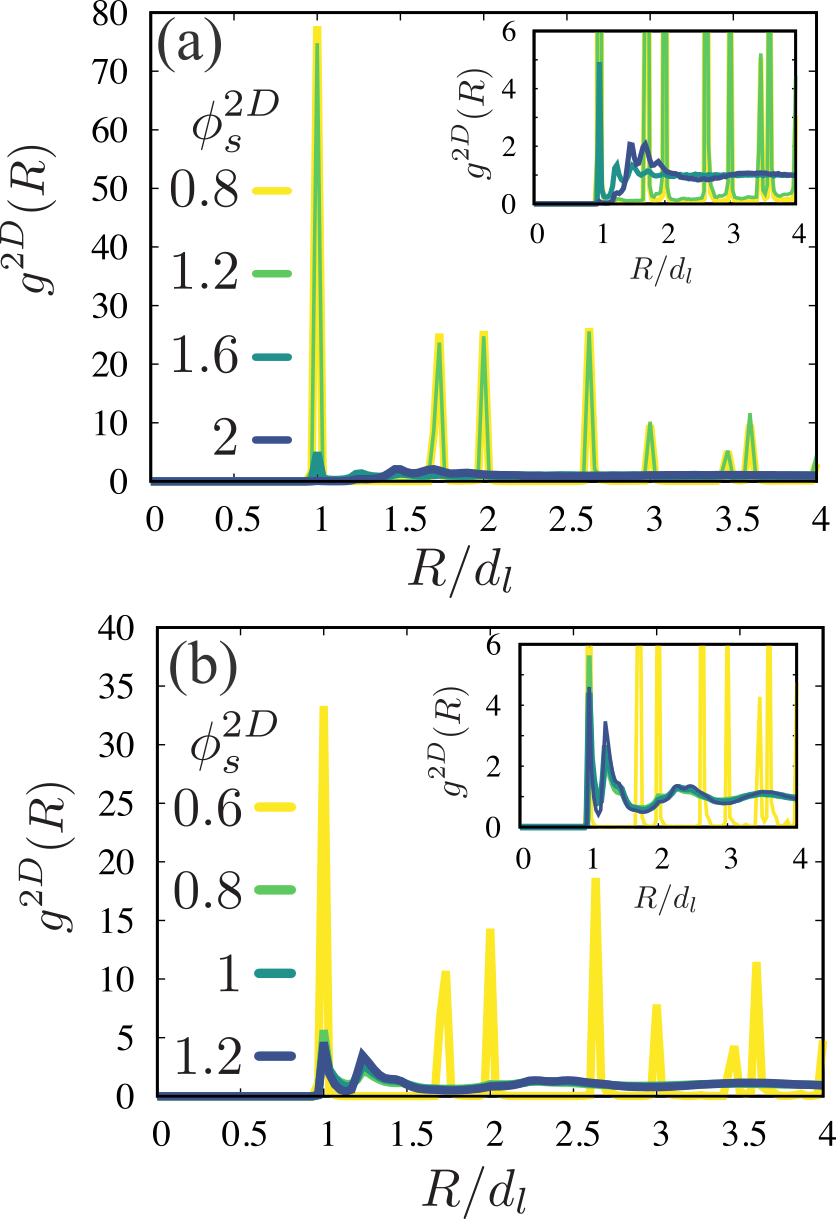}
    \caption{
    \fmrevised{
    Radial distribution function of large particles for various $\phi^{\text{2D}}_s$ with the size ratio of $4$ ($d_l=\SI{3}{mm}$ and $d_s=\SI{0.75}{mm}$).
    The representative cases, (a) $\phi^{\text{2D}}_l=0.2$ and (b) $\phi^{\text{2D}}_l=0.55$, are presented.
    The data are time-averaged within $\SI{300}{s}<t<\SI{600}{s}$.
    The insets show a magnified view of the vertical axis to enhance visibility.}}
    \label{fig:rdf_sizeratio}
\end{figure}

\begin{figure}
    \centering
    \includegraphics[width=0.8\textwidth]{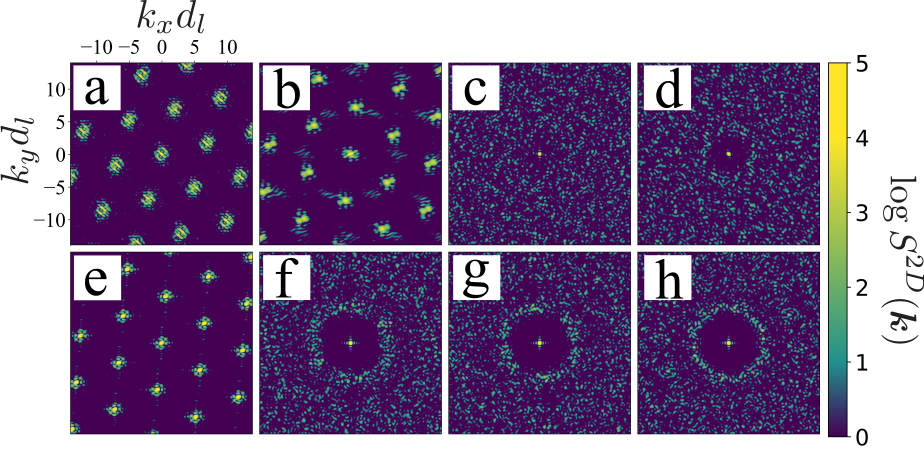}
    \caption{
    \fmrevised{
    Structure factor corresponding to Fig.~\ref{fig:snapshot_sizeratio}, as defined by Eq.~\eqref{eq:sk}, for various packing fractions of large and small granular particles: $\phi^{\text{2D}}_l$, $\phi^{\text{2D}}_s$ are (a)(0.2, 0.8), (b)(0.2, 1.2), (c)(0.2, 1.6), (d)(0.2, 2.0), (e)(0.55, 0.6), (f)(0.55, 0.8), (g)(0.55, 1), and (h)(0.55, 1.2).}}
    \label{fig:structure_factor_sizeratio}
\end{figure}

\fmrevised{
Following our analysis above, we now examine the effect of the size ratio of the particles.
To this end, we calculate the structure with the size ratio $4$ ($d_l=\SI{3}{\mm}$ and $d_s=\SI{0.75}{\mm}$).
The snapshots, radial distribution functions, and the structure factor are presented in Figs.~\ref{fig:snapshot_sizeratio}, ~\ref{fig:rdf_sizeratio}, and ~\ref{fig:structure_factor_sizeratio}, respectively.
From Fig.~\ref{fig:snapshot_sizeratio}, strong segregation clearly emerges when the small particle packing fraction is low for both $\phi_l^{\text{2D}}=0.2$ and $0.55$.
This strong segregation induces a local hexagonal structure, as shown in Fig.~\ref{fig:snapshot_sizeratio}(a), (b), and (e). This local hexagonal phase is also characterized by strong peaks in RDFs (Fig.~\ref{fig:rdf_sizeratio}) and hexagonal patterns in the structure factor as shown in Fig.~\ref{fig:structure_factor_sizeratio}(a), (b), and (e).
This segregation disappears as $\phi_s^{\text{2D}}$ increases, similar to the case of $d_l/d_s=3$. However, in the simulated density regime, the global hexagonal phase at large $\phi_s^{\text{2D}}$ is not observed.
From these calculations, reducing segregation appears to be insensitive to the size ratio, whereas the hexagonal order is sensitive to changes in the particle size ratio.
}

\begin{figure}
    \centering
    \includegraphics[width=0.8\textwidth]{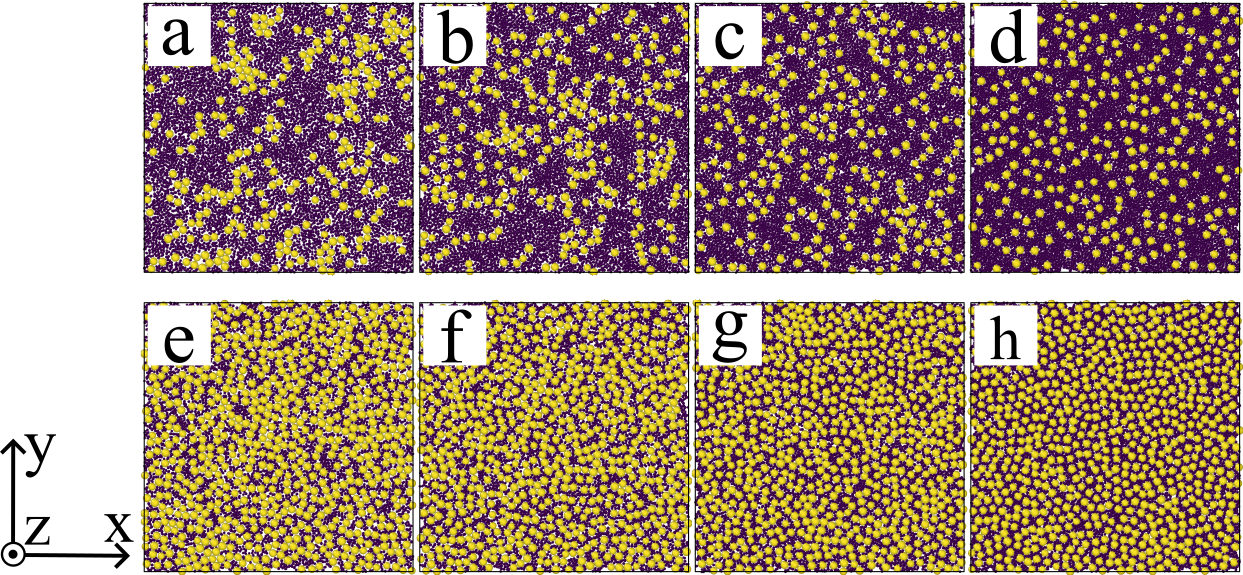}
    \caption{
    \fmrevised{
    Snapshots for various 2D packing fractions in a bidisperse case. Particle diameters are uniformly distributed within $d_l\pm 10\%$ and $d_s\pm 10\%$, where $d_l=\SI{3}{\mm}$ and $d_s=\SI{1}{\mm}$.
    The 2D packing fractions ($\phi^{\text{2D}}_l$, $\phi^{\text{2D}}_s$) are (a)(0.2, 0.8), (b)(0.2, 0.9), (c)(0.2, 1.0), (d)(0.2, 1.2), (e)(0.55, 0.45), (f)(0.55, 0.5), (g)(0.55, 0.6), and (h)(0.55, 0.7). All snapshots are obtained after the long-time vibration: $t=\SI{600}{\s}$.}}
    \label{fig:snapshot_bidisperse}
\end{figure}

\begin{figure}
    \centering
    \includegraphics[width=0.55\textwidth]{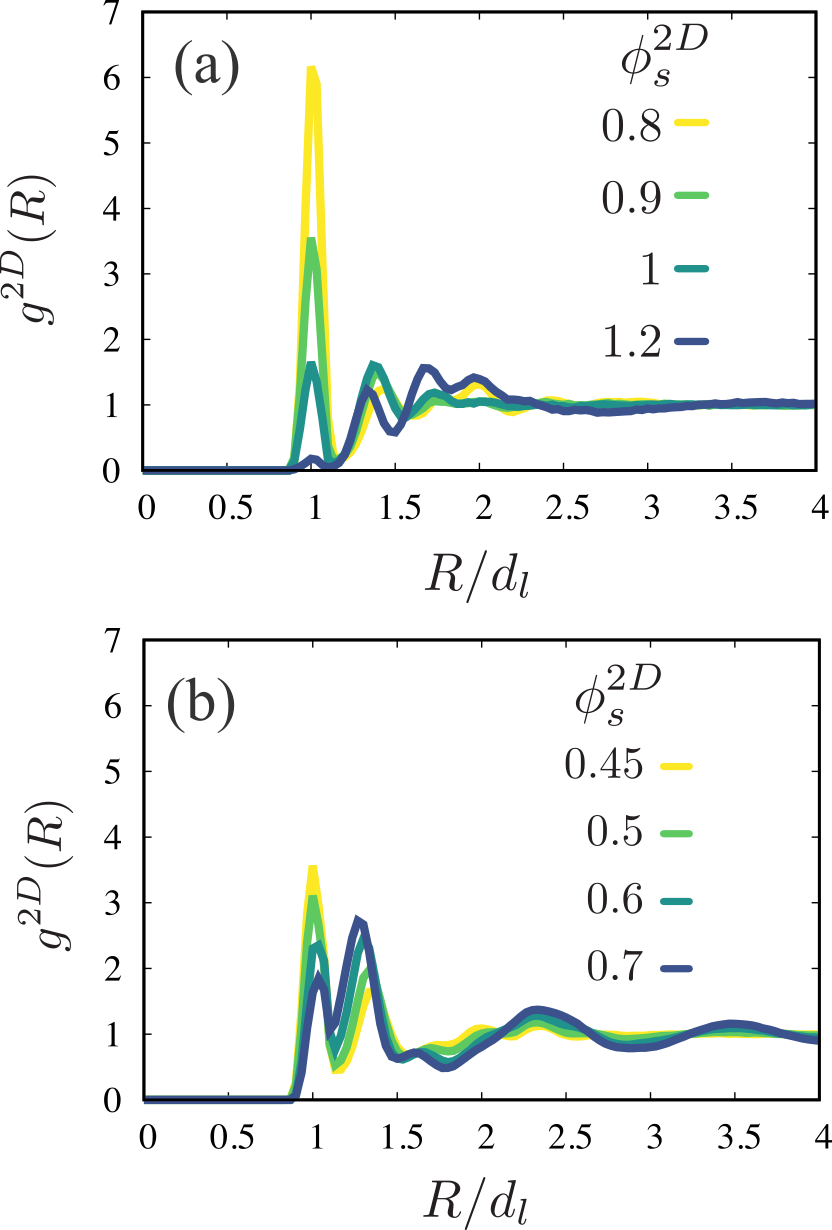}
    \caption{
    \fmrevised{
    Radial distribution function of large particles for various $\phi^{\text{2D}}_s$ in bidisperse cases. Particle diameters are uniformly distributed within $d_l\pm 10\%$ and $d_s\pm 10\%$, where $d_l=\SI{3}{\mm}$ and $d_s=\SI{1}{\mm}$.
    Representative cases are presented for (a) $\phi^{\text{2D}}_l=0.2$ and (b) $\phi^{\text{2D}}_l=0.55$.
    Data are time-averaged over main computation $\SI{300}{\second}<t<\SI{600}{\second}$.}}
    \label{fig:rdf_bidisperse}
\end{figure}

\begin{figure}
    \centering
    \includegraphics[width=0.8\textwidth]{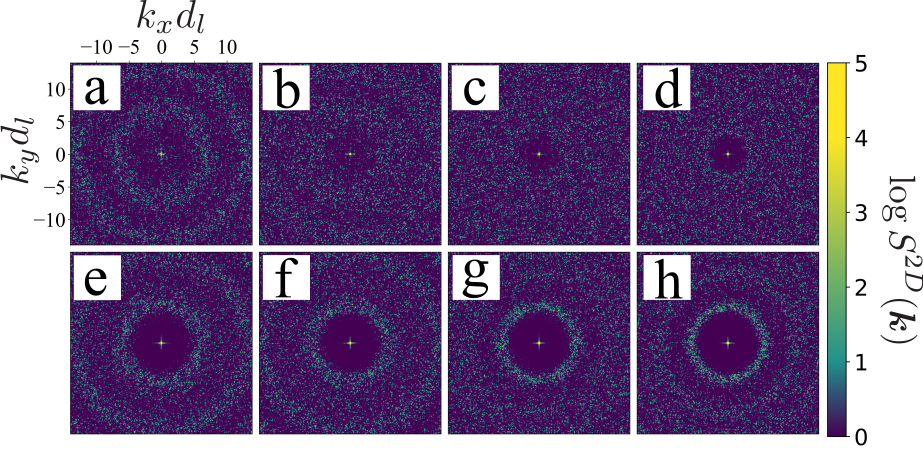}
    \caption{
    \fmrevised{
    Structure factors corresponding to Fig.~\ref{fig:snapshot_bidisperse}, as defined by Eq.~\eqref{eq:sk}, for various 2D packing fractions in bidisperse systems. The packing fractions ($\phi^{\text{2D}}_l$, $\phi^{\text{2D}}_s$) of large and small particles are (a)(0.2, 0.8), (b)(0.2, 0.9), (c)(0.2, 1.0), (d)(0.2, 1.2), (e)(0.55, 0.45), (f)(0.55, 0.5), (g)(0.55, 0.6), and (h)(0.55, 0.7).}}
    \label{fig:structure_factor_bidisperse}
\end{figure}

\fmrevised{The shape of the size distribution can affect the structure.
In research on glassy media, it is known that slight differences in particle size can inhibit crystallization \cite{kob1997dynamical}.
To examine the effect of the size distribution shape, we consider bidisperse cases with uniform size distributions for large and small particles within $d_l\pm 10\%$ and $d_s\pm 10\%$, where $d_l=\SI{3}{mm}$ and $d_s=\SI{1}{mm}$.
Snapshots, radial distribution functions, and structure factors obtained at time after long time vibrations ($t=\SI{600}{s}$) are shown in Figs.\ref{fig:snapshot_bidisperse}, \ref{fig:rdf_bidisperse}, and \ref{fig:structure_factor_bidisperse}, respectively.
From the snapshots and RDFs, we observe that segregation disappears as $\phi_s^{\text{2D}}$ increases for both $\phi_l^{\text{2D}}=0.2$ and $0.55$, which is consistent with the binary-sized case.
However, Fig.~\ref{fig:structure_factor_bidisperse} shows that the hexagonal phase, which emerges in the binary-sized setup at $\phi_l^{\text{2D}}=0.55$ (as shown in Fig.~\ref{fig:structure_factor}), is not observed in the bidisperse case.
Namely, while the reduction of segregation is insensitive to the variances in size distributions of large and small particles, the formation of hexagonal order is sensitive to these variances.
Additionally, this result indicates that the reduction of segregation occurs even in conditions where crystallization is inhibited, suggesting that the mechanism behind the segregation reduction does not stem from the symmetrical ordering of particles.}

\begin{figure}[htbp]
    \centering
    \includegraphics[width=1.\columnwidth]{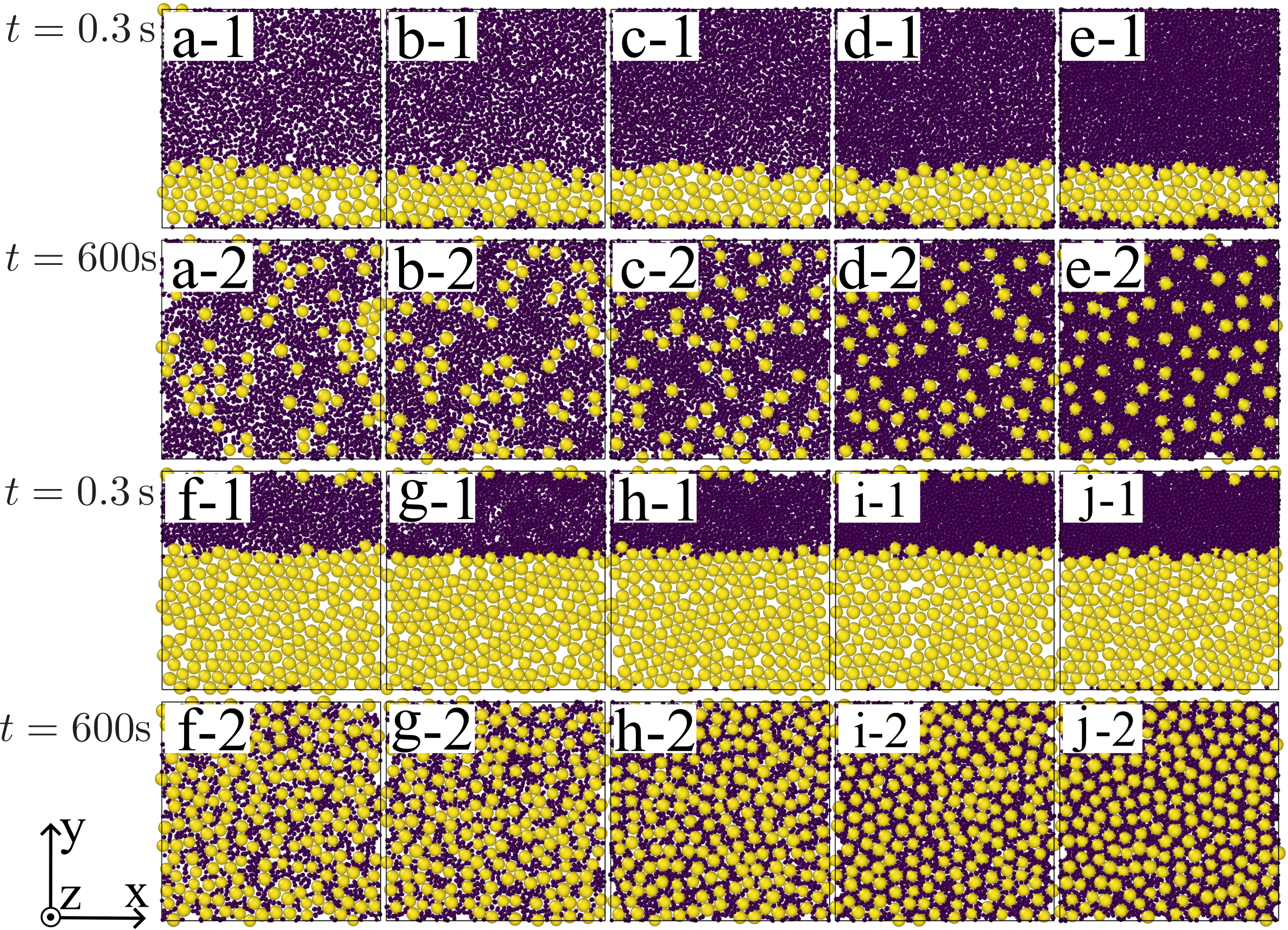}
    \caption{
    \fmrevised{
    Snapshots of the system initiated from separated configurations.
    The diameters of large particles are uniformly distributed between $0.9d_l$ and $1.1d_l$ with $d_l=\SI{3}{\mm}$, and those of small particles between $0.9d_s$ and $1.1d_s$ with $d_l=\SI{1}{\mm}$.
    The system size is $L=\SI{50}{\mm}$. The 2D packing fractions $\phi^{\text{2D}}_l$, $\phi^{\text{2D}}_s$ are set to be (a-1,2)(0.2, 0.8), (b-1,2)(0.2, 0.9), (c-1,2)(0.2, 1.0), (d-1,2)(0.2, 1.2), (e-1,2)(0.2, 1.4), (f-1,2)(0.55, 0.45), (g-1,2)(0.55, 0.5), (h-1,2)(0.55, 0.6), (i-1,2)(0.55, 0.7), and (j-1,2)(0.55, 0.8). Subscripts 1 and 2 denote snapshots taken immediately after vibration ($t=\SI{0.3}{\s}$) and after the long-time vibrations ($t=\SI{600}{\s}$), respectively. All snapshots are projected onto the $xy$-plane for clarity.
    \label{fig:snapshot_separated}}
    }
\end{figure}

\begin{figure}
    \centering
    \includegraphics[width=0.55\textwidth]{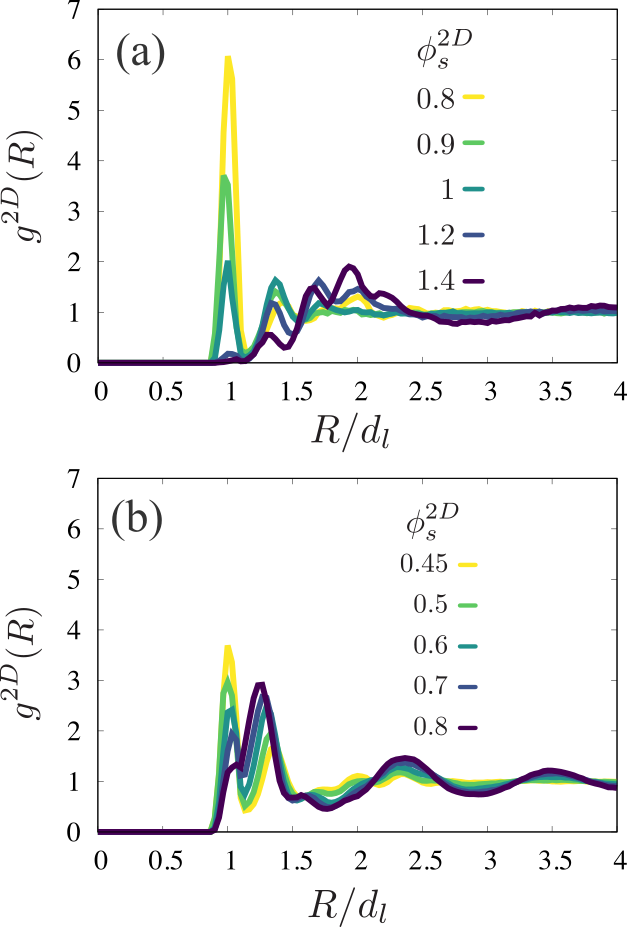}
    \caption{
    \fmrevised{Radial distribution functions of large particles averaged over the main calculation period $\SI{300}{\s}<t<\SI{600}{\s}$ initiated from the separated configurations shown in Fig.~\ref{fig:snapshot_separated}.
    The diameters of large particles are uniformly distributed between $0.9d_l$ and $1.1d_l$ with $d_l=\SI{3}{\mm}$, and those of small particles between $0.9d_s$ and $1.1d_s$ with $d_s=\SI{1}{\mm}$.
    The system size is $L=\SI{50}{\mm}$.
    The representative cases, (a) $\phi^{\text{2D}}_l=0.2$ and (b) $\phi^{\text{2D}}_l=0.55$, are presented with various $\phi^{\text{2D}}_s$.}}
    \label{fig:rdf_separated}
\end{figure}

\begin{figure}
    \centering
    \includegraphics[width=0.8\textwidth]{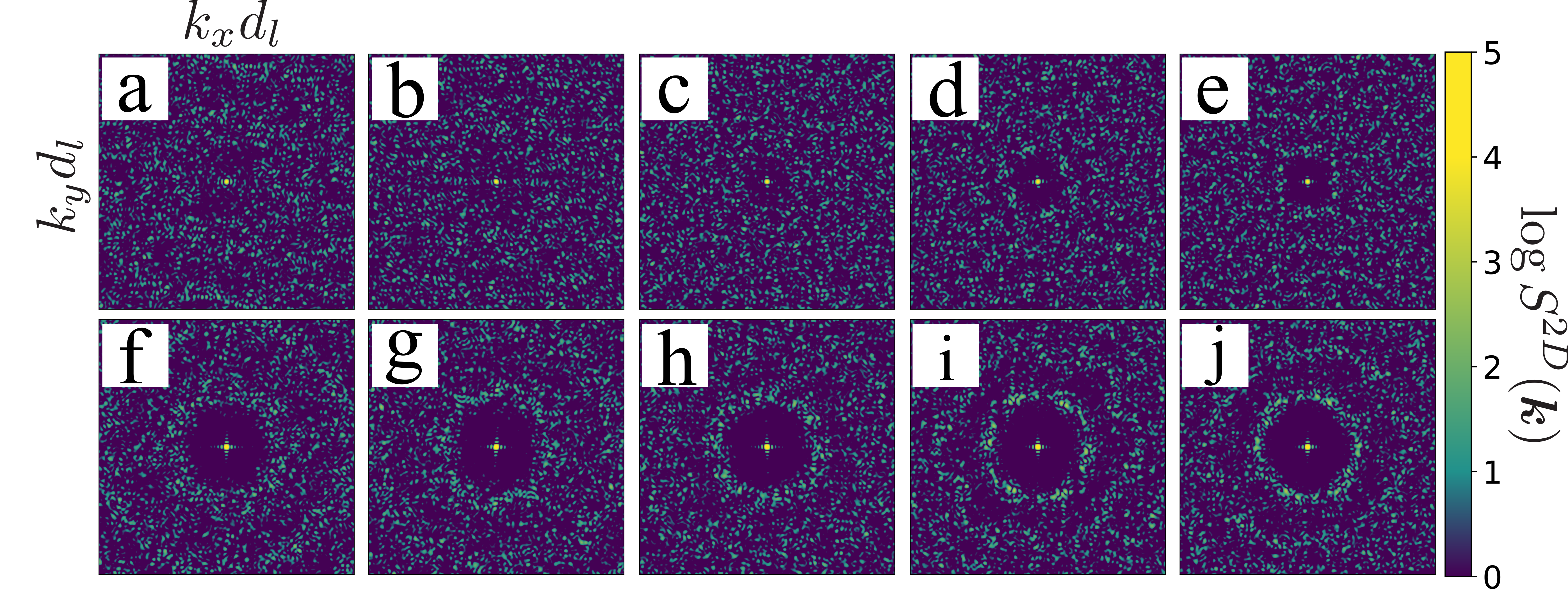}
    \caption{
    \fmrevised{
    Structure factors (Eq.~\eqref{eq:sk}) corresponding to Fig.~\ref{fig:snapshot_separated} after the long-time vibrations ($t=\SI{600}{s}$).
    The 2D packing fractions $\phi^{\text{2D}}_l$, $\phi^{\text{2D}}_s$ are set to be (a)(0.2, 0.8), (b)(0.2, 0.9), (c)(0.2, 1.0), (d)(0.2, 1.2), (e)(0.2, 1.4), (f)(0.55, 0.45), (g)(0.55, 0.5), (h)(0.55, 0.6), (i)(0.55, 0.7), and (j)(0.55, 0.8).
    }
    }
    \label{fig:structure_factor_separated}
\end{figure}

\fmrevised{
Before discussions, we should also check the effect of the initial configuration.
While we have started the simulation with random positions, as presented in Fig,~\ref{fig:snapshot}, we also perform the simulation from completely separated structures for bidisperse cases where the particle sizes are uniformly ranged $d_l\pm 10\%$ and $d_l\pm 10\%$ with $d_l=\SI{3}{mm}$ and $d_s=\SI{1}{mm}$.
Figs.~\ref{fig:snapshot_separated} shows the snapshots immediately after vibration and those after long-time vibrations ($t=\SI{600}{\s}$).
When the large particles are diluted $\phi^{2D}_l=0.2$ and $\phi^{2D}_s$ is small, the separated structure breaks with time, but the small clusters remain, which repeat creation and annihilation.
The small clusters disappear as  $\phi^{2D}_s$ increases.
A similar trend of the reduction of the segregation can be observed when the large particles are dense ($\phi^{2D}_l=0.8$), while the hexagonal structure does not appear due to the particle size dispersion.
The radial distribution functions and structure factors obtained after the long-time vibrations, initiated from the separated configurations, are presented in Fig.~\ref{fig:rdf_separated}, and \ref{fig:structure_factor_separated}.
The obtained results are almost the same as those initiated from random structures, as presented in Figs~\ref{fig:rdf_bidisperse} and \ref{fig:structure_factor_bidisperse}.
Namely, within the simulated parameter regime, the initial configuration does not affect the results after the long-time vibrations.
}

\section{Discussion}
When mixing granular materials of two different sizes, segregation generally appears \cite{rosato2020segregation}.
This behavior is observed even in the system confined by parallel walls \cite{melby2007depletion, galanis2010depletion}, which is the subject of this study. 
As shown in Figs.~\ref{fig:snapshot} and~\ref{fig:rdf} with small $\phi^{\text{2D}}_s$, the larger particles segregate, consistent with previous studies \cite{melby2007depletion, rivas2011segregation}.
Meanwhile, by thoroughly investigating the cases for $\phi^{\text{2D}}_s+\phi^{\text{2D}}_l\ge 1$, a regime that has not been studied, we discovered a phenomenon overlooked by prior research, as depicted in Figs.~\ref{fig:snapshot} and~\ref{fig:rdf}.
Namely, by introducing an excess of small particles, an effective repulsive force arises between the larger particles, which is longer than the diameter of the large particles $d_l$, eliminating the segregation.
This segregation reduction is confirmed for a different size ratio, system size, vibrational acceleration, and shape of size distribution.
Due to the effective repulsion, as shown in Figs.~\ref{fig:structure_factor} and~\ref{fig:order_parameter}, the larger particles can exhibit a hexagonal phase at a lower packing fraction than in a monocomponent system at the size ratio $3$.
The setup employed in this study is experimentally feasible, and thus the validation of our results would be straightforward.

The effective repulsion manifested by the excessive introduction of small particles might be akin to an entropic force in an equilibrium system.
The vibrated granular particles move randomly due to collisions with surrounding particles, and it may be plausible that the macroscopic structure is determined to maximize the volume in which each particle can move.
In a colloidal system consisting of large and small particles in an equilibrium state, the attractive force between large particles can emerge \cite{shah2003viscoelasticity, roth2000depletion, biben1996depletion}, which is known as the depletion interaction explained by the Asakura-Osawa model \cite{asakura1958interaction}.
This model assumes that the small particles are dilute and they are uniformly distributed in space.
When the small particles are not diluted, the small particles construct a complex liquid structure \cite{hansen2013theory}, and the conventional Asakura-Osawa model would not be straightforwardly applied.
Experimentally, it is reported that the entropic repulsive force between the large particles can appear when the surrounding smaller objects are not dilute \cite{crocker1999entropic}.
In the granular material, the entropy-like attractive force has been reported for the binary-sized mixture \cite{melby2007depletion} and particle-rod mixture \cite{galanis2006spontaneous}.
Thus, the entropy-like repulsive force between the larger particles might also be possible for the moderately dense case where the structure of the small particles is liquid-like.
Of course, since granular systems are not in an equilibrium state, there is no justification that the concept of entropy defined only in an equilibrium state can be applied. 
Additionally, granular systems can exhibit pattern formations due to differences in mass \cite{rivas2011segregation}, dissipation strength \cite{perera2010diffusivity}, and asymmetry in particle diffusivity \cite{weber2016}, which cannot be explained by equilibrium theories.
Although the observed repulsive behavior is similar to that in the colloidal system in an equilibrium state, it must be emphasized that we cannot determine that the mechanism of the effective repulsion between large granular particles is explained based on the entropic-like force.

Recent studies have shown intriguing results in a similar setting.
Plati et al. \cite{plati2024quasi} numerically investigated the structure of a binary-sized granular system confined between two parallel walls, focusing on regions with a two-dimensional packing fraction $\phi^{\text{2D}}_l+\phi^{\text{2D}}_s$ less than $1$.
The particle size ratio was set at $2$.
Upon vibrational excitation of the system, they discovered that the scattering pattern temporarily develops from a liquid-like state to a structure possessing eightfold rotational symmetry, indicative of a quasi-crystalline phase.
In this work, we studied the system with a particle size ratio of $3$ and investigated the regime with $\phi^{\text{2D}}_l+\phi^{\text{2D}}_s\ge 1$.
Within this simulated regime, we did not observe the manifestation of the quasi-crystalline scattering pattern.

Before the conclusion, it would be informative to discuss the application of this work.
This study examines the special setup: moderately dense granular particles confined between walls.
While such a setup is often employed in physical backgrounds \cite{reis2006crystallization, melby2007depletion, rivas2011segregation}, it may provide a fresh perspective on mixing processes in engineering.
Composite materials made by mixing particles into resins are significantly useful since the mixed particles can enhance some mechanical, thermal, and electrical properties of the materials \cite{christensen2012mechanics, clyne2019introduction}.
For instance, a composite material, which includes monolayered hollow particles, shows unique absorbing properties of electromagnetic waves \cite{qiao2020lattice, qiao2021preparation}.
Such a material property generally depends on the structure of the mixed particle \cite{roberts1996structure}, and controlling the mixing process is therefore important.
In literature \cite{qiao2020lattice}, the samples are prepared by fixing particles with a lattice pattern on the epoxy resin, subsequently pouring the rest of the resin, and performing thermal curing.
Our findings of the reduction of segregation and the emergence of the hexagonal phase possibly contribute to the dispersion techniques of such monolayered particles in the composite.

\section{Conclusion}

This study investigated the structures of the binary granular mixtures confined within parallel walls subjected to vertical vibration using the discrete element method.
The segregation of the large particle, which often arises in binary-sized mixtures, disappears upon increasing the small particle fraction.
From RDFs, the effective repulsive interaction between the large particles emerges for large $\phi^{\text{2D}}_s$.
The reduction in segregation by the emergence of the effective repulsive force is confirmed against a different system size, particle size ratio, vibrational acceleration, and even for the bidisperse case.
Further, at the size ratio of $3$, the structure factor and the hexagonal order parameter clearly indicate that the large particles exhibit the hexagonal phase below the fraction smaller than that of the monocomponent case.
The system settings can be experimentally realized, and we hope our findings will be confirmed.
The theoretical analyses may be rather difficult, but the interesting future work.
Furthermore, investigating the effects of contact properties such as rolling and twisting frictions, as well as attractive forces, on granular structure would be crucial for future work.
Our study provides fresh insight into the non-equilibrium statistical physics of the granular material.
Additionally, our work may contribute to the mixing and transport processes in chemical, food, and pharmaceutical engineering.

\section*{Acknowledgement}
\fmrevised{This work is inspired by the discussion during our participation in the long-term workshop "Frontier in Non-equilibrium Physics 2024" (YITP-T-24-01). We would like to acknowledge the warm hospitality during our stay in the YITP.}
F.N. was supported by Grant-in-Aid for JSPS (Japan
Society for the Promotion of Science) Fellows (Grant No.
JP24KJ0156).
K.Y. appreciates the financial support from JSPS Research Fellow (Grant No.21J13720) and Leave a Nest Co.,ltd.

\section*{Declaration}
There are no conflicts to declare.

\appendix
\renewcommand\thefigure{\thesection.\arabic{figure}}
\setcounter{figure}{0}

\section{Additional data}

\begin{figure}
    \centering
    \includegraphics[width=0.8\textwidth]{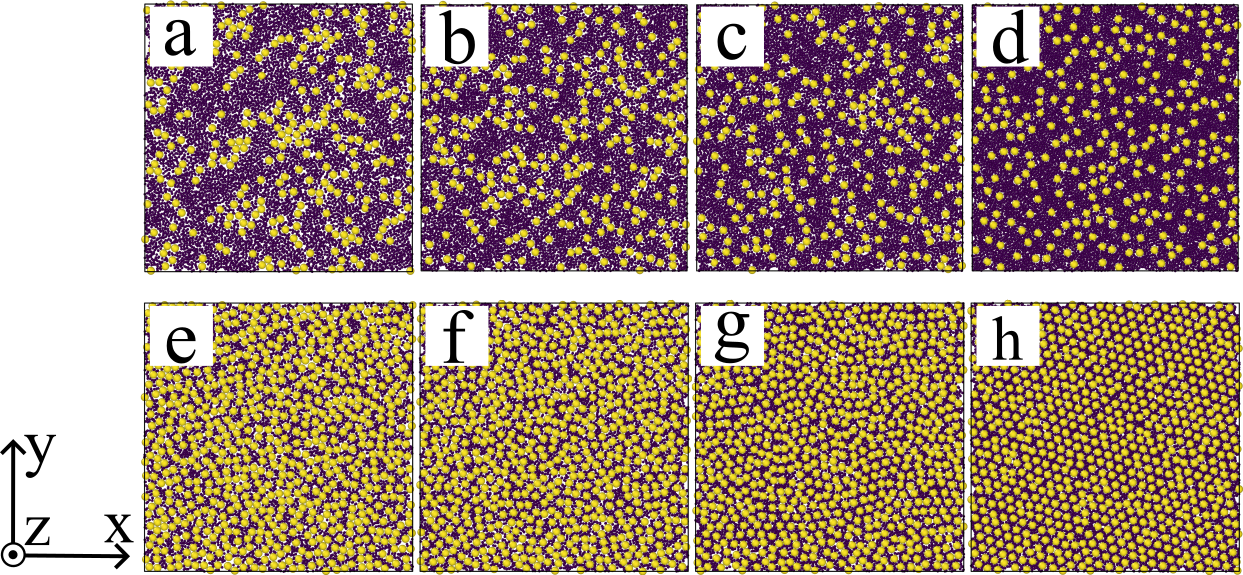}
    \caption{
    Snapshots of the system with a large angular frequency of vibration, $\Omega=\SI{2810}{\radian\per\s}$ ($A\Omega^2/g\simeq 250$).
    The particle diameters are $d_l=\SI{3}{\mm}$ and $d_s=\SI{1}{\mm}$, and the system length is $L=\SI{100}{\mm}$.
    The 2D packing fractions $\phi^{\text{2D}}_l$, $\phi^{\text{2D}}_s$ are (a)(0.2, 0.8), (b)(0.2, 0.9), (c)(0.2, 1.0), (d)(0.2, 1.2), (e)(0.55, 0.45), (f)(0.55, 0.5), (g)(0.55, 0.6), and (h)(0.55, 0.7). All snapshots are acquired after the long-time vibrations $t=\SI{600}{\s}$.}
    \label{fig:snapshot_acceleration}
\end{figure}

\begin{figure}
    \centering
    \includegraphics[width=0.55\textwidth]{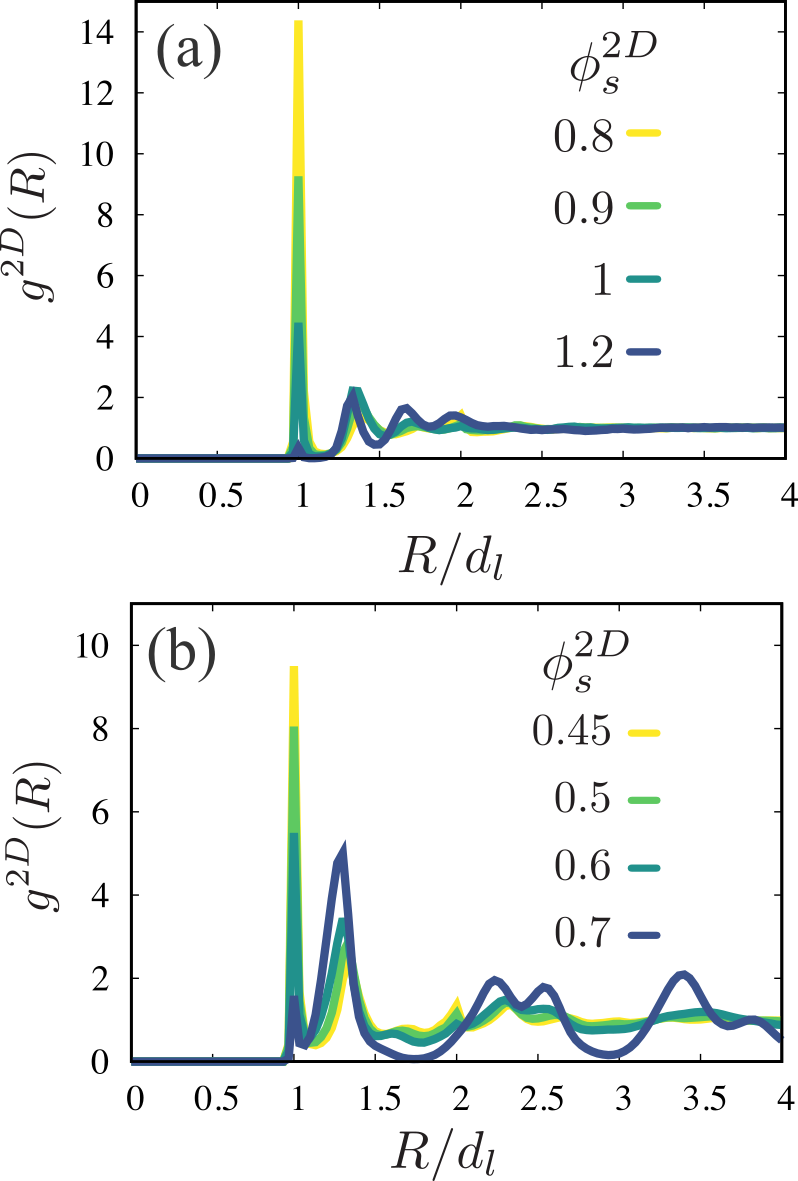}
    \caption{Radial distribution function of large particles for various $\phi^{\text{2D}}_s$ at a large angular frequency of vibration, $\Omega=\SI{2810}{\radian\per\s}$ ($A\Omega^2/g\simeq 250$).
    The particle diameters are $d_l=\SI{3}{\mm}$ and $d_s=\SI{1}{\mm}$.
    The representative cases, (a) $\phi^{\text{2D}}_l=0.2$ and (b) $\phi^{\text{2D}}_l=0.55$, are presented.
    Data are time-averaged within the main calculation period $\SI{300}{\s}<t<\SI{600}{\s}$.}
    \label{fig:rdf_acceleration}
\end{figure}

\begin{figure}
    \centering
    \includegraphics[width=0.8\textwidth]{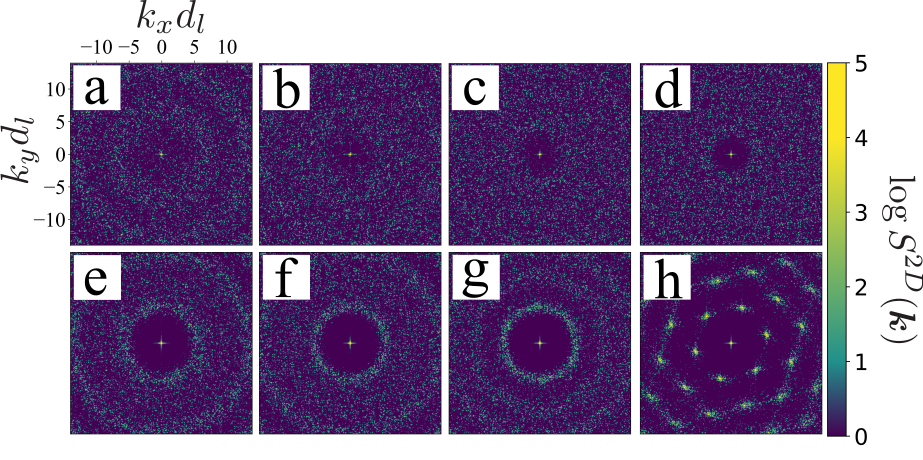}
    \caption{ 
    Structure factors corresponding to Fig.~\ref{fig:snapshot_acceleration}, as defined by Eq.~\eqref{eq:sk}, for various 2D packing fractions of large and small particles: ($\phi^{\text{2D}}_l$, $\phi^{\text{2D}}_s$) are (a)(0.2, 0.8), (b)(0.2, 0.9), (c)(0.2, 1.0), (d)(0.2, 1.2), (e)(0.55, 0.45), (f)(0.55, 0.5), (g)(0.55, 0.6), and (h)(0.55, 0.7).}
    \label{fig:structure_factor_acceleration}
\end{figure}

\begin{figure}
    \centering
    \includegraphics[width=0.8\textwidth]{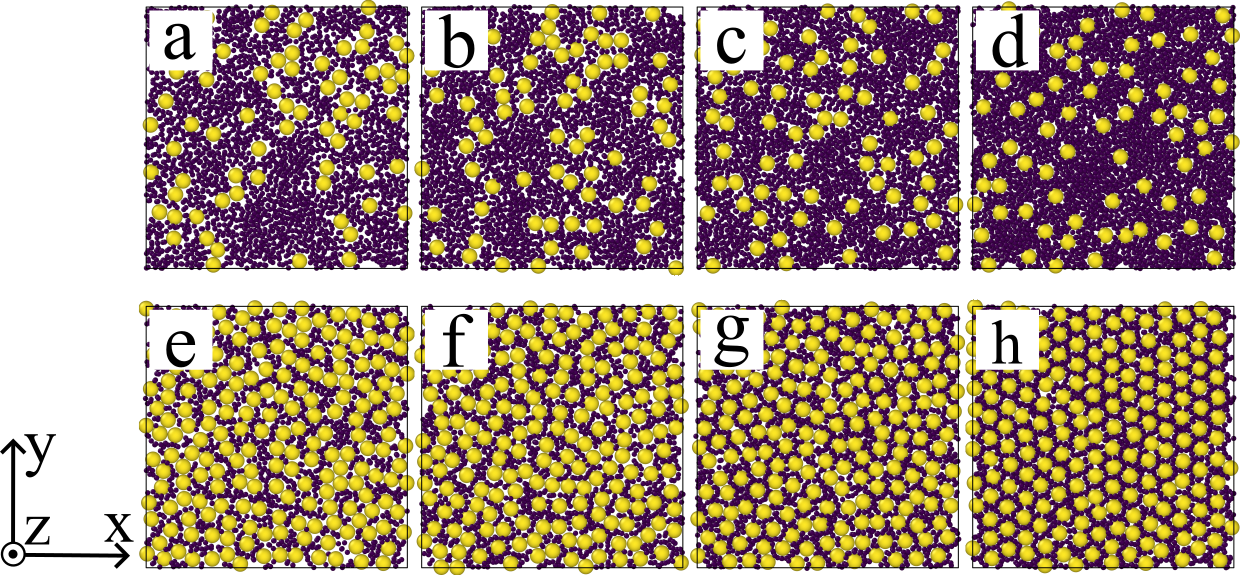}
    \caption{Snapshots of systems with various 2D packing fractions for a small system size ($L=\SI{50}{mm}$).
    The particle diameters are $d_l=\SI{3}{\mm}$ and $d_s=\SI{1}{\mm}$.
    The 2D packing fractions ($\phi^{\text{2D}}_l$, $\phi^{\text{2D}}_s$) are (a)(0.2, 0.8), (b)(0.2, 0.9), (c)(0.2, 1.0), (d)(0.2, 1.2), (e)(0.55, 0.45), (f)(0.55, 0.5), (g)(0.55, 0.6), and (h)(0.55, 0.7). All snapshots are acquired after the long-time vibrations $t=\SI{600}{\s}$.}
    \label{fig:snapshot_systemsize}
\end{figure}

\begin{figure}
    \centering
    \includegraphics[width=0.55\textwidth]{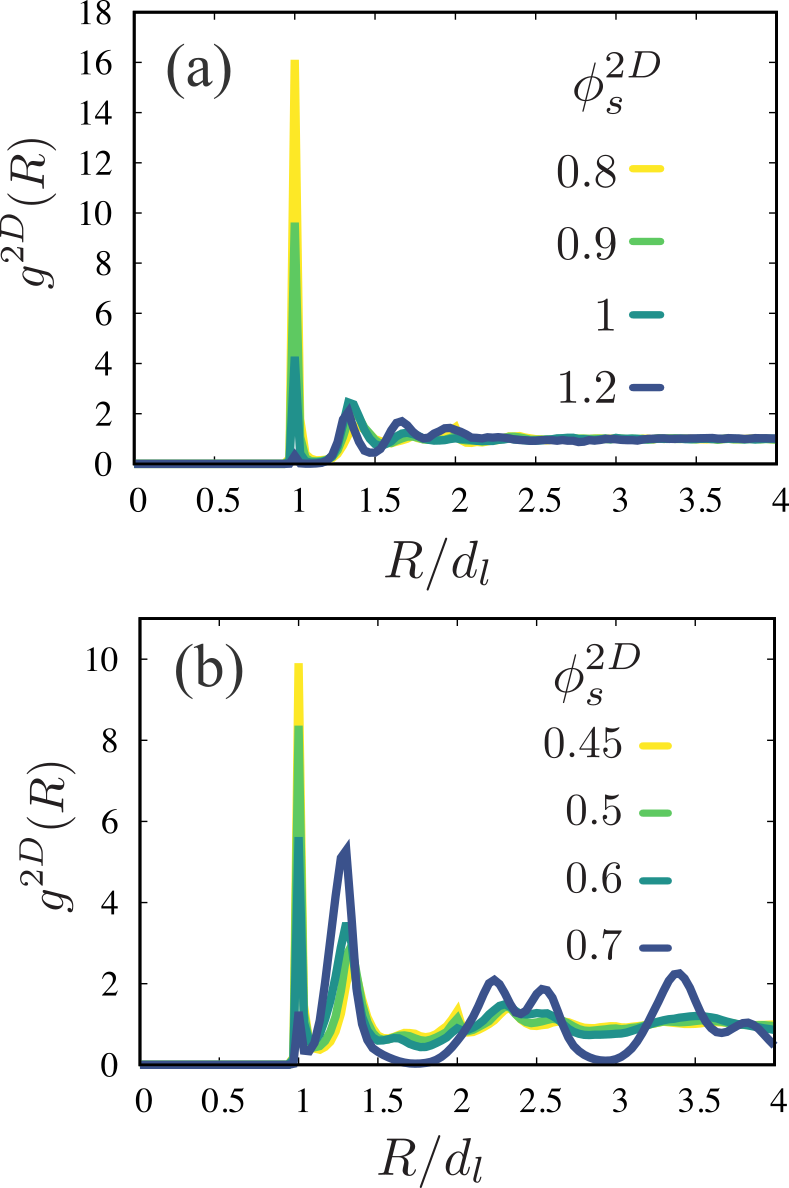}
    \caption{Radial distribution function of large particles for various $\phi^{\text{2D}}_s$ in a small system with $L=\SI{50}{\mm}$.
    The particle diameters are $d_l=\SI{3}{\mm}$ and $d_s=\SI{1}{\mm}$.
    The representative cases, (a) $\phi^{\text{2D}}_l=0.2$ and (b) $\phi^{\text{2D}}_l=0.55$, are presented.
    Data are time-averaged over the main calculation period $\SI{300}{\s}<t<\SI{600}{\s}$.}
    \label{fig:rdf_systemsize}
\end{figure}

\begin{figure}
    \centering
    \includegraphics[width=0.8\textwidth]{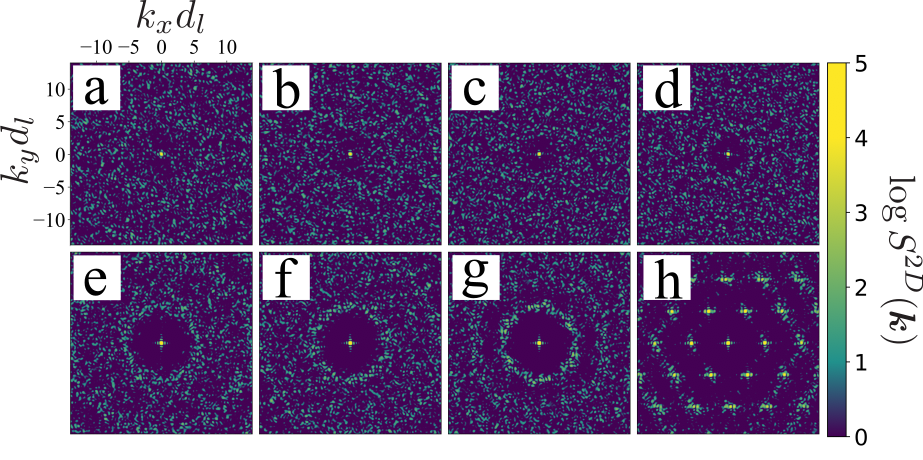}
    \caption{
    Structure factors corresponding to Fig.~\ref{fig:snapshot_systemsize}, as defined by Eq.~\eqref{eq:sk}, for various 2D packing fractions of large and small particles in the small system $L=\SI{50}{\mm}$. The packing fractions ($\phi^{\text{2D}}_l$, $\phi^{\text{2D}}_s$) are (a)(0.2, 0.8), (b)(0.2, 0.9), (c)(0.2, 1.0), (d)(0.2, 1.2), (e)(0.55, 0.45), (f)(0.55, 0.5), (g)(0.55, 0.6), and (h)(0.55, 0.7).}
    \label{fig:structure_factor_systemsize}
\end{figure}

Figs~\ref{fig:snapshot_acceleration}, \ref{fig:rdf_acceleration}, and \ref{fig:structure_factor_acceleration} display the snapshots, radial distribution functions, and structure factors with various $\phi^{\text{2D}}_l$ and $\phi^{\text{2D}}_s$, at a large angular frequency: $\Omega=2810$ ($A\Omega^2/g\simeq 250$).
Figs~\ref{fig:snapshot_systemsize}, \ref{fig:rdf_systemsize}, and \ref{fig:structure_factor_systemsize} present the snapshots, radial distribution functions, and structure factors for a small system size: $L=\SI{50}{mm}$.

\bibliography{ref.bib}

\end{document}